\documentstyle[12pt]{article}
\begin{document}
\newcommand{\bea}{\begin{eqnarray}}
\newcommand{\eea}{\end{eqnarray}}
\newcommand{\be}{\begin{equation}}
\newcommand{\ee}{\end{equation}}
\newcommand{\non}{\nonumber}
\newcommand{\ov}{\overline}
\global\parskip 6pt
\begin{titlepage}
\begin{center}
{\Large\bf Topological Modes in Dual Lattice Models}\\
\vskip 1.0in
Mark Rakowski \footnote{Email: Rakowski@maths.tcd.ie}   \\
\vskip .10in
{\em Dublin Institute for Advanced Studies, 10 Burlington Road, }\\
{\em Dublin 4, Ireland}\\
\end{center}
\vskip .10in
\begin{abstract}
  Lattice gauge theory with gauge group $Z_{P}$ is reconsidered in
four dimensions on a simplicial complex $K$. One finds that the dual theory,
formulated on the dual block complex $\hat{K}$, contains topological modes
which are in correspondence with the cohomology group $H^{2}(\hat{K},Z_{P})$,
in addition to the usual dynamical link variables. This is a general
phenomenon in all models with single plaquette based actions; the
action of the dual theory becomes twisted with a field representing
the above cohomology class. A similar observation is made about the dual
version of the three dimensional Ising model. The importance of distinct
topological sectors is confirmed numerically in the two dimensional
Ising model where they are parameterized by $H^{1}(\hat{K},Z_{2})$.
\end{abstract}
\vskip 1.0in
\begin{center}
\end{center}

\begin{center}
October 1994\\
DIAS Preprint 94-32
\end{center}
\end{titlepage}

\section{Introduction}

   The use of duality transformations in statistical systems has
a long history, beginning with applications to the two dimensional
Ising model \cite{KW}. Here one finds that the high and low temperature
properties of the theory are related.
This transformation has been extended to many
other discrete models and is particularly useful when the symmetries
involved are abelian; see \cite{RS} for an extensive review. All
these studies have been confined to hypercubic lattices, or other
regular structures, and these have rather limited global topological
features.

   Since lattice models  are defined in a way which depends clearly
on the connectivity of links or other regions, one expects some
sort of topological effects generally. In this paper we will reconsider
four dimensional $Z_{P}$ lattice gauge theory based on actions which
are functions of the holonomy around single plaquettes. The underlying
lattice will be a simplicial complex which we take to model spacetime.
To such a simplicial complex there is naturally associated a dual
block complex which maps each $k$-simplex  to a ``dual block''
of codimension $k$. This
parallels the standard construction for hypercubic lattices only now
the whole construction can be applied to spacetimes with more subtle
topology.
We find that the dual of $Z_{P}$ lattice gauge theory in four
dimensions, is another lattice gauge theory on the dual block complex,
only now the link based dynamical variables are twisted with extra
topological modes parameterized by the second cohomology group
of the dual block complex with coefficients in $Z_{P}$.
One might regard this as a
coupling of normal lattice gauge theory  with a topological field theory.

   We begin by reviewing some important material on simplicial complexes,
paying particular attention to some subtleties of the dual block
construction. The dual transformation is then applied to four dimensional
gauge theory as well as the Ising model in three dimensions. We also
show explicitly on a two dimensional torus that the topological
sectors in the dual Ising model are generally distinct.

\section{Simplices and Dual Blocks}

   Here we will collect some standard definitions involving simplicial
complexes and their dual block complexes; we refer to \cite{JM} for
a complete exposition.

    Let $V=\{ v_{1},...,v_{N_{0}} \}$ be a collection of $N_{0}$
elements which we will call the {\em vertex} {\em set}. A {\em
simplicial complex} $K$ is a collection of finite nonempty subsets
of $V$ such that if $\sigma \in K$ so is every nonempty subset of
$\sigma$. An element of $K$ is called a {\em simplex} and its {\em
dimension} is one less than the number of vertices it contains. We
picture the 1-dimensional simplex $\{ v_{i}, v_{j} \}$ as the line
segment connecting two distinct points in Euclidean space.
Similarly, $\{ v_{i},v_{j},v_{k} \}$ can be pictured as the triangle
with the three indicated vertices. An {\em orientation} of a simplex
$\{ v_{0},...,v_{m} \}$, denoted $[ v_{0},...,v_{m} ]$ is an
equivalence class of the ordering of the vertices according to even
and odd permutations; this gives a direction to line segments and
a direction of circulation around the vertices of a triangle etc.

   Let $K^{(m)}=\{ \sigma_{\alpha} \}$ denote the collection of all oriented
$m$-simplices in $K$. The group of $m-chains$ on $K$, denoted by
$C_{m}(K)$,
is defined to be the set of all finite linear combinations $\sum_{\alpha}
\, n_{\alpha}\,\sigma_{\alpha}$ with integer coefficients. A {\em boundary}
operator $\partial_{m}: C_{m}(K)\rightarrow C_{m-1}(K)$ is defined by:
\be
\partial_{m} [v_{0},...,v_{m}] = \sum_{i=0}^{m} \; (-1)^{i} \;
[v_{0},...,\bar{v}_{i},...,v_{m}] \;\;
\ee
where we omit the vertex corresponding to $\bar{v}_{i}$. The $m-cochains$
of $K$ with coefficients in the abelian group G is $C^{m}(K,G) =
Hom(C_{m}(K),G)$, and we have a {\em coboundary} operator $\delta^{m}:
C^{m}(K,G)\rightarrow C^{m+1}(K,G)$ defined through the evaluation
of an $m$-cochain $c$ on an $(m+1)$-simplex $\sigma$:
\be
< \delta c, \sigma > = < c , \partial \sigma > \;\; .
\ee
Let $Z^{m}(K,G)$ denote the kernel of $\delta^{m}$ and $B^{m}(K,G)$
the image of $\delta^{m-1}$; the {\em cohomology group} $H^{m}(K,G)$
is then defined as the quotient $Z^{m}/ B^{m}$.

   We will be using the dual block complex associated to a given
simplicial complex and this is defined in terms of a subdivision
of the original complex. Given a geometric realization of a simplex,
the central point is called the {\em barycenter}. The {\em  barycentric
subdivision} $Sd(K)$ of $K$ is the new simplicial complex obtained
by subdividing every simplex in $K$ at its barycenter. In this case,
a 1-simplex becomes a union of two 1-simplices. Similarly a 2-simplex
becomes divided into six 2-simplices; the new vertices are at the center
of the original simplex as well as the center of each bounding
1-simplex. Given the simplicial complex $K$, the simplices of $Sd(K)$
are of the form $\pm [\hat{\sigma}_{i_{1}},... \hat{\sigma}_{i_{m}} ]$,
where $dim(\sigma_{i_{1}}) > \cdots > dim(\sigma_{i_{m}})$, and
$\hat{\sigma}$ denotes the barycenter of $\sigma$.
Associated to an $m$-simplex
$\sigma$ of $K$, we define the {\em dual block} $D(\sigma)$ to be the
union of all open simplices of $Sd(K)$ for which $\hat{\sigma}$ is
the final vertex.
If $K$ satisfies the additional technical conditions to be an
$n$-dimensional PL-manifold \cite{Ro}, which we will assume, then it
consists entirely of $n$-simplices and
their faces, and
the closed dual block, denoted by $\bar{D}(\sigma)$, has dimension
$n-m$. One also has $\partial \bar{D}(\sigma) =
\bar{D}(\sigma) - D(\sigma)$.
The {\em dual block complex} $\hat{K}$ is the collection of all blocks
which are dual to the simplices of $K$.

   Let us discuss some examples to make the general setting a bit more
concrete. Take the simplicial complex which represents $S^{4}$ given
by the 4-simplices on the boundary of the 5-simplex $[0,1,2,3,4,5]$.
Here the complex contains, 6 vertices, 15 links, 20 2-simplices, and
15 3-simplices in addition to the 6 4-simplices. This case is exceptional
in that the dual block complex is also a simplicial complex in its
own right. It is not surprising since duality exchanges vertices
for 4-blocks and links with 3-blocks and the number of simplices of these
types match.

A more interesting example is given by a triangulation
of $CP^{2}$ \cite{Ku}. This complex has nine vertices, and
is fully determined by
specifying the 4-simplices which are 36 in number and these are listed
in \cite{Ku}.
In this case, the dual block complex
is not a simplicial complex and here it is useful to represent a given block
by the blocks which lie on its boundary.
To form the dual block complex, we begin by associating a 0-block
(dual vertex) to each of the 4-simplices in $K$.
The 1-blocks correspond to 4-simplices which share a common
3-face. The 2-blocks are slightly more
difficult to enumerate, but one can
make use of a theorem \cite{JM} which
states that $\partial \bar{D}(\sigma)$ {\em is the
union of all blocks} $D(\tau)$ {\em for which} $\tau$ {\em has}
$\sigma$ {\em as a proper face}.
In this particular complex, 2-blocks contain from three to six
dual links on their boundary. $\hat{K}$ is clearly not a simplicial complex
but both $K$ and $\hat{K}$ have subdivisions in common and encode
the same topological information. One can proceed to enumerate all
blocks with the help of the above theorem.

\section{Dual Transformation}

  The partition function of $Z_{P}$ lattice gauge theory is defined as
a sum over all link variables $ U_{ij}\in Z_{P} $ which we represent
multiplicatively, and a Boltzmann weight factor for every 2-simplex
in the simplicial complex $K$ \cite{MC},
\be
Z = \sum_{\{ U_{ij} \} } \;
\prod_{ \Delta \in K^{(2)} } \; \exp [ S(U_{\Delta}) ] \;\;
.\label{z}
\ee
The action $S$ is a function of the holonomy $U_{\Delta}$ around $\Delta$,
\be
U_{\Delta} = U_{ij}\, U_{jk}\, U_{ki}\;\; ,
\ee
where the boundary of $\Delta$ is given by,
\be
\partial \Delta = [j,k] - [i,k] + [i,j] \;\; .
\ee
The character expansion of the Boltzmann weight is
\bea
\exp[S(U)] = \sum_{n=0}^{P-1} \; b_{n} \; U^{n} \;\; ,
\eea
where the $b_{n}$ coefficients are the parameters of the theory. Usually
one requires that the Boltzmann weight be insensitive to the orientations
of the holonomies and this will restrict $b_{P-n} = b_{n}$.
Introducing an integer $n_{\Delta}\in \{ 0,...,P-1 \}$
for each 2-simplex $\Delta$, $Z$ becomes,
\be
Z= \sum_{\{ U_{ij} \} }
\; \prod_{\Delta \in K^{(2)}} \;
\sum_{\{ n_{\Delta} \}} \; b_{n_{\Delta}} \; U_{\Delta}^{n_{\Delta}}
\;\; .
\ee
The collection of the $n_{\Delta}$ for all 2-simplices in $K$ is a
2-cochain.
Now rearrange the order of factors, the idea being to collect all
terms proportional to each link variable $U_{ij}$; we have
\bea
Z= \sum_{\{ n_{\Delta} \}}\; \prod_{\Delta \in K^{(2)}}
b_{n_{\Delta}} \; \prod_{ [i,j]\in K^{(1)} } \;
( \sum_{ U_{ij} } \; (
\prod_{\Delta \supset [i,j]}\; U_{ij}^{n_{\Delta}} ))\;\; ,
\eea
where the last product in this equation is over all 2-simplices which
contain the specified link $[i,j]$.
Using the representation of a mod-$P$ delta function,
\be
\sum_{U\in Z_{P}} \; U^{n} = P \; \delta(n) \;\; ,
\ee
one obtains,
\bea
Z= P^{N_{1}}\; \sum_{\{ n_{\Delta} \}}\; \prod_{\Delta \in K^{(2)}}
b_{n_{\Delta}} \; \prod_{  [i,j]\in K^{(1)} } \;
 \delta(\sum_{\Delta \supset [i,j]}\; n_{\Delta})\;\; .    \label{zdelta}
\eea
Notice that the sum in the delta function is over all 2-simplices
which contain $[i,j]$ as a face.
Now let us map each of the above quantities into the dual picture.
First, each of the integer variables $n_{\Delta}$ associated to
the 2-simplex $\Delta$ in $K^{(2)}$ becomes associated with the
unique closed 2-block $\bar{D}(\Delta)$ in four dimensions;
\be
n_{\bar{D}(\Delta)} = n_{\Delta} \;\; .
\ee
The utility of the
dual transformation becomes apparent when one looks at the assembly
of delta functions in (\ref{zdelta}). Each of these is associated to
a link $[i,j]$ of the original simplicial complex; in the dual picture
there is one delta function for each 3-block $\bar{D}([i,j])$.
Moreover, one has that
\be
\sum_{\Delta \supset [i,j]} \; n_{\Delta} =
\sum_{\bar{D}({\Delta}) \subset \partial \bar{D}([i,j])}
n_{\bar{D}(\Delta)}
\;\; .
\ee
To see this, we can appeal to Theorem 64.1 of \cite{JM} which states
that $\partial \bar{D}(\sigma)$ {\em is the union of all blocks} $D(\tau)$
{\em for which} $\tau$ {\em has} $\sigma$ {\em as a proper face}; in
this case we take $\sigma = [i,j]$. So far, the above presentation
parallels the usual case of a hypercubic lattice \cite{RS,MC}, but
now we see that we have the possibility of non-trivial solutions
to the constraints.
The last delta function says that the sum of the 2-cochains around
the boundary of each 3-block must vanish mod-$P$; these are
nothing but the conditions for a 2-cocycle with coefficients in
$Z_{P}$ on the dual block complex. To solve those constraints, we
need the kernel of the coboundary operator,
\be
\delta^{2} : C^{2}(\hat{K},Z_{P}) \rightarrow C^{3}(\hat{K},Z_{P})
\ee
operating on the 2-cochains of $\hat{K}$. The partition function
expressed in terms of dual quantities is then
\be
Z = P^{\hat{N}_{3}} \sum_{ \{ n_{\hat{\Delta}}\} \in
Z^{2}(\hat{K},Z_{P}) } \;\;
\prod_{\hat{\Delta}\in \hat{K}^{(2)}} b_{n_{\hat{\Delta}}} \;\; ,\label{zd}
\ee
where $\hat{K}^{(2)}$ is the set of all 2-blocks and
$\hat{N}_{m}= N_{4-m}$ is the number of $m$-blocks.
$Z^{2}(\hat{K},Z_{P})$ is nothing but the
image of $\delta^{1}$, the trivial 2-cocycles, together with the
non-trivial cocycles which represent the cohomology classes:
\be
Ker(\delta^{2}) = Im(\delta^{1}) \oplus H^{2}(\hat{K},Z_{P}) \;\; .
\ee
Let us parameterize the solution by:
\be
n_{\hat{\Delta}} = B_{\hat{\Delta}} + (\delta^{1} A)_{
\hat{\Delta}} \;\; ,
\ee
where $B$ runs over the 2-cochains which cannot be written as the
boundary of a link field as in the second term. Here $A$ and $B$
take their values in the additive group $\{ 0, ... , P-1 \}$.
We would like to write $Z$ as a sum over $A$ and $B$ but we need to
take account of the fact that $\delta^{1}$ has a kernel; a simple
sum over $A$ would be an overcounting.
To go a bit further, let
us restrict the following discussion to the case where $P$ is
a prime number so
$Z_{P}$ is an algebraic field. The kernel of $\delta^{1}$, the
group of 1-cocycles, is parameterized
by the image of $\delta^{0}$ together with $H^{1}(\hat{K},Z_{P})$.
When $P$ is prime,
the later cohomology group is then a sum of copies of $Z_{P}$; let
$h^{1}$ denote the number of these copies, i.e. the dimension of
$H^{1}(\hat{K},Z_{P})$ as a vector space over $Z_{P}$. The dimension
of $Ker(\delta^{1})$ is then,
\bea
dim\;\; Ker(\delta^{1}) &=& h^{1} + dim \;\; Im(\delta^{0}) \\
&=& h^{1}   + (\hat{N}_{0} - 1) \;\; .\non
\eea
The expression for Z becomes,
\be
Z = P^{(\hat{N}_{3}-\hat{N}_{0} - h^{1} + 1)}
\sum_{ B \in H^{2}(\hat{K},Z_{P}) }\; \sum_{ \{ A_{\hat{i}\hat{j}}
\} }\;
\prod_{\hat{\Delta}\in \hat{K}^{(2)}} b_{(B+\delta^{1} A)_{\hat{\Delta}}}
\;\; .
\ee
In the dual theory, the new Boltzmann weight is therefore just proportional
to $b_{(B+\delta^{1} A)}$. In multiplicative $Z_{p}$ notation,
\be
U_{\hat{i}\hat{j}} = \exp[ \frac{2 \pi i}{P} \; A_{\hat{i}\hat{j}}]
\;\;\; , \;\;\; W_{\hat{\Delta}} = \exp[\frac{2 \pi i}{P} \;
B_{\hat{\Delta}}] \;\; ,
\ee
and the dual action is then a function of the product
\be
W_{\hat{\Delta}} \; U_{\hat{\Delta}} \;\; .
\ee
Let us define the dual Boltzmann weight by,
\be
\exp[\hat{S}(W_{\hat{\Delta}} \; U_{\hat{\Delta}})] =
P^{(\hat{N}_{3}-\hat{N}_{0} - h^{1} + 1)}
\; b_{(B+\delta^{1} A)_{\hat{\Delta}}} \;\; ,
\ee
and we have the final form of the partition function for four
dimensional $Z_{P}$ gauge theory, only now
written in terms of the dual complex variables,
\be
Z =
\sum_{ B \in H^{2}(\hat{K},Z_{P}) }\; \sum_{ \{ U_{\hat{i}\hat{j}}
\} }\;
\prod_{\hat{\Delta}\in \hat{K}^{(2)}}
\exp[\hat{S}(e^{\frac{2 \pi i}{P} B_{\hat{\Delta}}}\; U_{\hat{\Delta}})]
\;\; .
\ee
We see then that the dual theory is generally another link based
gauge theory with an action that depends on the usual holonomy of
link fields but now twisted with an extra phase. This phase is
topological in origin and one can consider the $B$ field to be
a topological excitation of a discrete topological field theory
\cite{BR,BR2,BR3}.
Since any lattice which captures the same topology will have the same
number of these modes in the dual theory, it would be interesting
to understand their role in the continuum limit. Both the original
theory on $K$ and the dual formulation on $\hat{K}$ have common
subdivisions and in taking a continuum limit of $K$ one is inducing
a continuum limit in the dual picture. The extra modes captured in
$H^{2}(\hat{K},Z_{P})$ are topological and, in a sense, already
associated with the continuum limit.
Perhaps the simplest
example to study where these modes are present is $CP^{2}$ where
$H^{2}(CP^{2},Z_{P}) = Z_{P}$. We have remarked that this
has a very simple simplicial presentation \cite{Ku} and would be
suitable for numerical studies.

   It should not be surprising that the duality between the
Ising model and $Z_{2}$ lattice gauge theory in three dimensions
has the same extra set of topological modes. Consider the Ising model
defined on a simplicial complex $K$ where the partition
function is given by
\be
Z_{I} = \sum_{\{ s_{i} \}} \; \prod_{ [i,j]\in K^{(1)} } \exp[
\beta\, s_{i} s_{j} ] \;\; .   \label{ising}
\ee
Here one will make a character expansion by introducing an integer
variable $n_{ij}\in \{ 0,1\}$ for each link. Following the same
procedure as before, one finds that the dual variables in three
dimensions,
\be
n_{\hat{\Delta}} = n_{\bar{D}([i,j])} = n_{ij}
\ee
are restricted to satisfy the 2-cocycle condition, and the 3d partition
function $Z_{I3}$ becomes identical in structure to (\ref{zd}),
\be
Z_{I3} = 2^{\hat{N}_{3}} \sum_{ \{ n_{\hat{\Delta}}\} \in
Z^{2}(\hat{K},Z_{2}) } \;\;
\prod_{\hat{\Delta}\in \hat{K}^{(2)}} b_{n_{\hat{\Delta}}} \;\; .
\ee
The subsequent steps are the same as in the previous case.
Here we can be more explicit since
we began with a definite action, and the final expression
can be written as,
\bea
Z_{I3} = 2^{(\hat{N}_{3}-\hat{N}_{0}-h^{1}+1)}
e^{\hat{\beta}_{0}\hat{N}_{2}}
\sum_{ B \in H^{2}(\hat{K},Z_{2}) }\;
\sum_{\{ U_{\hat{i}\hat{j}} \} }\;
\prod_{\hat{\Delta} \in \hat{K}^{(2)}}
\exp[ \hat{\beta}\, (-1)^{ B_{\hat{\Delta}} }\, U_{\hat{\Delta}} ]\;\;,
\eea
with the new parameters $\hat{\beta}$ and $\hat{\beta}_{0}$ given by,
\bea
\hat{\beta} &=& \frac{1}{2}\; \ln[\coth \beta]  \label{par}\\
\hat{\beta}_{0} &=& \frac{1}{2}\; \ln[\cosh \beta \cdot \sinh \beta]
\;\; . \non
\eea
The dual theory is then a three
dimensional $Z_{2}$ lattice gauge theory coupled with extra
topological modes associated to $H^{2}(\hat{K},Z_{2})$. A simple
space where these modes will appear is $RP^{3}$ which has
$H^{2}(RP^{3},Z_{2}) = Z_{2}$.

   Similarly, an analysis of the two dimensional Potts model will
yield a dual theory which depends on a topological mode
belonging to $H^{1}(K,Z_{P})$. In this case, one can see these
modes on the torus with a square lattice. Let us look now at the
two dimensional Ising model specified by the partition function
(\ref{ising}). A square lattice with opposite sides identified
has a dual lattice of precisely the same structure, and the number
of vertices, links and squares is preserved by duality. In the
dual picture we have in general,
\be
Z_{I2} = 2^{\hat{N}_{2}-1} \sum_{ B \in H^{1}(\hat{K},Z_{2})}\;
\sum_{\{ A_{i} \} }\;
\prod_{[\hat{i},\hat{j}]} b_{(B+\delta^{0}A)_{\hat{i}\hat{j}}} \;\; .
\ee
The factor of $2^{-1}$ arises from the fact that $dim\, Ker (\delta^{0})
= 1$. Since the square lattice for $T^{2}$ and its dual are identical, the
partition function can be written as,
\be
Z_{I2} = 2^{N_{0}-1} e^{\hat{\beta}_{0}N_{1}}\;
\sum_{ B \in H^{1}(K,Z_{2})}\;
\sum_{\{ s_{i} \} }\;
\prod_{[i,j]} \; \exp[ \hat{\beta}\, (-1)^{B_{ij}}\, s_{i}\, s_{j} ]\;\;,
\label{isd}
\ee
with the new parameters $\hat{\beta}$ and $\hat{\beta}_{0}$ as in
equation (\ref{par}).
The change in the parameters in the dual theory is the same as
for the infinite square lattice \cite{RS,MC}; here the topology
of the torus has been accounted for.
A representation of the cohomology class $H^{1}(T^{2},Z_{2}) =
Z_{2}\oplus Z_{2}$ is easy to find. This amounts to specifying
the value of $B$ on all links. Perhaps the simplest choice is
to take $B$ to be zero on all links except that the vertical links
in the bottom row of squares are assigned $x$, and the horizontal
links in the right column of squares are assigned $y$. The sum
over $H^{1}$ amounts then to summing $x$ and $y$ over $\{ 0,1\}$.
Equation (\ref{isd}) has been checked numerically and one finds that
the contributions from the topological sectors generally differ from
each other.

The topological sectors in the dual two dimensional Ising model
are presumably related to spin structures (boundary conditions)
in the continuum limit, and this
supports the view that they will be important in the other
models we have considered.
Similar sectors have been observed from an entirely different
point of view \cite{X} in the correlation functions of the Ising
model on a cylinder.

   Let us just remark that in a similar way, lattice models based upon
the abelian group $U(1)$ will generally have topological modes in the
dual theory which are parameterized by cohomology classes with
integer coefficients.

{\bf Acknowledgements}\\
I am grateful to Michael Creutz and Siddhartha Sen for discussions.

\end{document}